\documentclass[twocolumn,showpacs,preprintnumbers,amsmath,amssymb,groupedaddress,prl]{revtex4}

\usepackage{graphicx}
\usepackage{dcolumn}
\usepackage{bm}
\usepackage{txfonts}
\newcommand{\ket}[1]{\mbox{$\vert #1 \rangle$}}

\newcommand{\Rb}{\mbox{$^{87}$Rb}}

\begin{document}


\title{Coherent collisional spin dynamics in optical lattices}
\author{Artur Widera}\email{Widera@Uni-Mainz.DE}
\author{Fabrice Gerbier}
\author{Simon F\"olling}
\author{Tatjana Gericke}
\author{Olaf Mandel}
\author{Immanuel Bloch}
\affiliation{Johannes Gutenberg-Universit\"at, Staudingerweg 7,
55099 Mainz, Germany}
\date{\today}

\begin{abstract}
We report on the observation of coherent, purely collisionally driven spin dynamics of neutral atoms in an optical lattice. For high lattice depths, atom pairs confined to the same lattice site show weakly damped Rabi-type oscillations between two-particle Zeeman states of equal magnetization, induced by spin changing collisions. This paves the way towards the efficient creation of robust entangled atom pairs in an optical lattice. Moreover, measurement of the oscillation frequency allows for precise determination of the spin-changing collisional coupling strengths, which are directly related to fundamental scattering lengths describing interatomic collisions at ultracold temperatures.
\end{abstract}

\maketitle
The creation and manipulation of spinor Bose-Einstein condensates (BEC) in optical traps \cite{StamperKurnReview01,Barrett01,Schmaljohann04,Chang04,Kuwamoto04,Higbie05} has opened a wide field of fascinating phenomena originating from the spin degree of freedom. Spinor systems allow to explore quantum magnetism \cite{Auerbach98} in a range of parameters not easily accessible in solid state physics. They have been proposed e.~g.~for the investigation of spin waves \cite{McGuirk02,Gu04} or for the study of strongly correlated ground states in optical lattices (see \cite{Demler02,Paredes03,Imambekov03,GarciaRipoll04} and references therein). More importantly, new aspects have emerged, such as the generation of entangled states \cite{Pu00,Duan00}, or dynamical spin mixing induced by interatomic collisions \cite{Ho98,Ohmi98,Law98,Pu99,Ciobanu00,Ueda02}. The latter phenomenon has been investigated in several recent experiments with Bose-Einstein condensates in optical dipole traps \cite{Schmaljohann04,Chang04,Kuwamoto04}. The fundamental mechanism responsible for many of these spin phenomena is a coherent collisional process in which the spin of each colliding particle is changed while the total magnetization is preserved.

In this Letter, we investigate this microscopic collision process in an ensemble of isolated atom pairs localized to lattice sites of a deep optical lattice. We observe coherent oscillations between two-particle Zeeman states, coupled by the spin-changing interaction. We show that for a broad range of parameters this dynamics can be described by a Rabi-type model. Our system allows for a precise measurement of the coupling strength for spin changing collisions. 
Finally, we observe an increasing damping of the oscillations with decreasing lattice depth, which we attribute to the onset of tunneling in the system.

\begin{figure}[htbp]
	\begin{center}
		\includegraphics[scale=0.65]{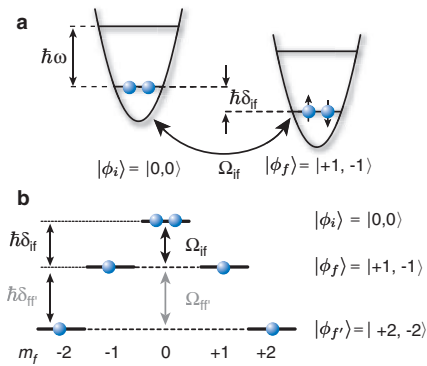}
	\end{center}
	\caption{(a) Two atoms localized in the vibrational ground state of a common lattice well can change their spin orientation while preserving total magnetization. The atoms remain in the lowest vibrational state at all times. (b) The process can be described as a coherent coupling between two-particle Zeeman states \ket{\phi_i}, \ket{\phi_f} and \ket{\phi_{f^\prime}}, where the coupling strength $\Omega_{if}$ depends on the choice of initial and final state, and the detunings $\delta_{if}$, $\delta_{ff^\prime}$ can be varied by the second order Zeeman shift.}
	\label{fig:Intro_v4}
\end{figure}

Let us consider a pair of \Rb\ atoms, localized in the vibrational ground state of a deep trapping potential, as shown in \mbox{Fig.~\ref{fig:Intro_v4}a}. The trapping frequency $\omega$ is assumed to be much larger than the typical interaction energy, so that excitations into higher vibrational levels are suppressed. The atom pair can then be solely described by a spin wavefunction $\ket{f_1,m_1; f_2,m_2}$, where $f_i, m_i$ are the total angular momentum and its projection onto the $z$-axis, and $i=1,2$ labels the first and second atom, respectively. In the following, we assume that both atoms are in the upper hyperfine ground state with $f_1=f_2=2$, and abbreviate the non-symmetrized two particle states as $\ket{m_1,m_2}$. In the absence of an external radio-frequency (rf) field, interatomic collisions drive the spin evolution of this system. 
For two interacting alkali atoms, the projection of the total angular momentum on the quantization axis is conserved, even in a finite magnetic field \cite{StamperKurnReview01,Barrett01,Schmaljohann04,Ho98,Ohmi98,Law98,Pu99}.
The interaction thus couples an initial state $\ket{\phi_i}\equiv\ket{m_1,m_2}$ to a final state $\ket{\phi_f} \equiv \ket{m_3,m_4}$, provided the total magnetization is conserved, i.~e.~$m_1+m_2 = m_3+m_4$. Furthermore, $s$-wave collisions between spin $f=2$ bosons are characterized by three scattering lengths $a_F$ for the collision channels with total spin $F$ ($F=0,2,4$) \cite{Ho98,Ueda02,Ciobanu00}. 
The matrix element $\Omega_{if}$ of the interaction hamiltonian between states \ket{\phi_i} and \ket{\phi_f} ("coupling strength") is proportional to $(4 \pi \hbar \Delta a)/M \; \int \vert \phi_0 \vert^4 d^3{\bf r}$, where $\phi_0$ is the (spin-independent) spatial wave function of the ground state in the potential well, $M$ is the atomic mass, and $\Delta a$ is a weighted difference of the scattering lengths $a_F$ that depends on the specific values of the magnetic quantum numbers. For example, in the case $\ket{\phi_i}=\ket{0,0}$ and $\ket{\phi_f} = \ket{+1,-1}$, one finds $\Delta a=(-7 a_0 - 5 a_2 + 12 a_4)/35$. For \Rb\ the different $a_F$ are almost equal (see e.~g.~\cite{Ciobanu00,vanKempen02}), so that the differences $\Delta a$ are typically only a few percent of the bare scattering lengths.

Due to the localization in the vibrational ground state and due to the constraint of a conserved magnetization, a given initial state can couple only to a few final two-particle states  participating in the spin evolution (see Fig.~\ref{fig:Intro_v4}b). In the case where only one final state is available (relevant for most experiments described below), the dynamics reduces to a Rabi-like model. 
We stress, however, that this system is quite different from the "usual" Rabi model coupling two \emph{single-particle} states under the influence of an external field. Here we investigate the coherent coupling between two \emph{two-particle} states induced by direct atomic interactions.
The Rabi model is parameterized by a coupling strength $\Omega_{if}$ discussed above, and a detuning $\delta_{if} = \delta_{0} + \delta(B^2)$ between the initial and final states, where $B$ is the value of the static external magnetic field. As the magnetization is conserved, the initial and final states experience the same first order Zeeman shift which has no influence on the spin dynamics. However, the second order Zeeman shifts are different and introduce the $B^2$-dependent detuning. The constant detuning $\delta_{0}$ originates from the difference in interaction energies in the initial and final states.
The atom pair thus is expected to oscillate between initial and final state at the effective Rabi frequency $\Omega^\prime_{if}= \sqrt{\Omega_{if}^2+\delta_{if}^2}$. 

In our experiment we investigate an array of localized atom pairs. We prepare a Mott-insulator of around $2\times10^5$ \Rb\ atoms in \ket{f=1,m_f=-1} in a combined optical and magnetic trap similar to our previous work \cite{Fölling05}.
In this work, unless stated otherwise, we use a lattice depth of $40\,E_r$. Here $E_r = h^2/2 M \lambda^2$ is the single photon recoil energy, and $\lambda \approx 840$\,nm the lattice laser wavelength. The overall harmonic confinement of the system leads to the creation of Mott-shells with different filling factors (see \cite{Fölling05} and references therein). For our parameters, we calculate that approximately half of the atoms are in the central region with two atoms per site. The remaining atoms are distributed in a surrounding shell with one atom per site and do not undergo atomic collisions. The number of sites with more than two atoms is negligible.

To study spin dynamics, we subsequently switch the magnetic trap off and load the sample into a pure optical lattice \cite{Widera05pre}. In order to preserve spin polarization of the atoms, a homogeneous magnetic field of approximately $1.2$\,G is maintained \cite{Note01}. The spin dynamics is initialized by transferring the sample into either $\ket{f=2,m_f=0}$ or $\ket{f=2, m_f=-1}$, and the magnetic field is subsequently ramped to a final value between $0.2$\,G  and $2$\,G. After time evolution for a variable time $t$ the optical trap is switched off. In order to spatially separate the different magnetic substates, a magnetic gradient field is switched on during the first 3\,ms of time-of-flight (TOF) \cite{StamperKurnReview01,Barrett01,Schmaljohann04,Chang04,Kuwamoto04}. The population $N_{m_j}$ of each Zeeman level $m_j$ is then detected after 7\,ms TOF with absorption imaging. 

\begin{figure}[htbp]
	\begin{center}
		\includegraphics[scale=0.7]{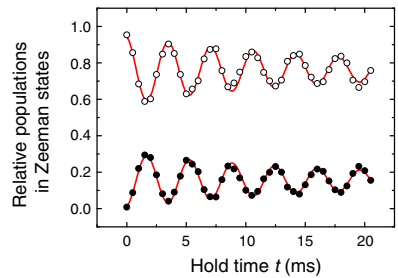}
	\end{center}
	\caption{Spin dynamics of atom pairs localized in an optical lattice at a magnetic field of $B=0.8$\,G. The atoms are initially prepared in $\ket{0,0}$ and can evolve into $\ket{+1,-1}$. Shown are the populations in $m_f=0$ ($\medcirc$) and $m_f=\pm 1$ ($\medbullet$) together with a fit to a damped sineyielding an oscillation frequency of $\Omega_{if}^\prime  = 2 \pi \times 278(3)$\,Hz.}
	\label{fig:Paper_Oscillations}
\end{figure}

We first consider the case where we start with both atoms in $\ket{\phi_i}=\ket{0,0}$. This state couples to  $\ket{\phi_f}=\ket{+1,-1}$ and $\ket{\phi_{f^\prime}}=\ket{+2,-2}$. The coupling strength for $\ket{\phi_i} \leftrightarrow \ket{\phi_{f^\prime}}$ is calculated to be two orders of magnitude smaller than for $\ket{\phi_i} \leftrightarrow \ket{\phi_f}$, and can be neglected. 
However, a two-step coupling channel $\ket{\phi_i} \leftrightarrow \ket{\phi_f} \leftrightarrow \ket{\phi_{f^\prime}}$ is also possible, with comparable coupling strengths for each step. Although present at low magnetic field \cite{Schmaljohann04,Kuwamoto04}, this two-step process is increasingly suppressed as the magnetic field is increased due to the large detuning. For $B> 0.6$\,G, the system mostly oscillates between $\ket{0,0}$ and $\ket{+1,-1}$.
This is shown in Fig.~\ref{fig:Paper_Oscillations} for the case of $B=0.8$\,G.  The relative populations in Fig.~\ref{fig:Paper_Oscillations} have been calculated as $N_0/N_{tot}$ for $\ket{0,0}$ and $(N_{+1}+N_{-1})/N_{tot}$ for $\ket{+1,-1}$, where $N_{tot}$ is the independently counted total atom number \cite{Note03}.
In order to describe the oscillations we use a Rabi-like model, where the transition probability to the final state is \cite{CohenTannoudji}
\begin{equation} \label{Eq:Rabi}
	P_f = \frac{\Omega_{if}^2}{\Omega_{if}^{\prime 2}}\; \frac{1}{2}\;\left( 1 - \cos \left( \Omega_{if}^\prime \, t \right) \, e^{-\gamma_{if} \,t} \right),
\end{equation}
with $\gamma_{if}$ being a phenomenological damping rate. The measured population in $m_f=\pm 1$ is simply given by $n\;P_f$, where $n$ is the fraction of atoms localized in doubly occupied lattice sites. A fit to the data using Eq.~\ref{Eq:Rabi} allows to extract the oscillation frequency and the damping rate. As explained above, for magnetic fields $B\leq0.6\,$G, the process $\ket{+1,-1} \leftrightarrow \ket{+2,-2}$ starts to play a role. In this magnetic field range we observe deviations from the pure sinusoidal evolution, as expected for a three level system. Due to limited space, the three level model will be presented elsewhere \cite{Widera05pre}.  

We have also investigated the case where the sample is prepared in the initial state $\ket{\phi} = \ket{-1,-1}$. The situation is simpler than in the previous case, because no third level exists to which the final state \ket{0,-2} could couple to while preserving magnetization. In this case the description as a two level system is appropriate at any magnetic field.

\begin{figure}[htbp]
	\begin{center}
		\includegraphics[scale=0.7]{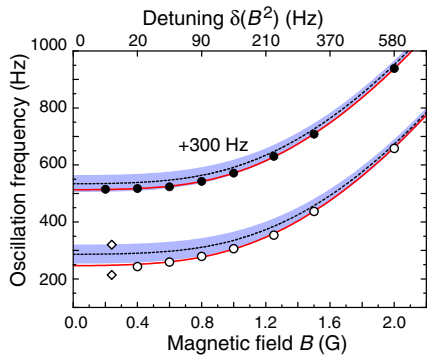}
	\end{center}
	\caption{Oscillation frequency of spin dynamics vs.~magnetic field (corresponding detuning in upper axis) for the case $\ket{0,0} \leftrightarrow \ket{+1,-1}$ ($\medcirc$) and $\ket{-1,-1}\leftrightarrow \ket{0,-2}$ ($\medbullet$). The diamonds are results of a Fourier transform for the coupled three-level case. The solid lines are fits to the expected behaviour of the effective Rabi frequency. The upper curve has been offset by 300\,Hz for clarity. The error bars are typically on the order of a few percent. As comparison the calculated curves (based on \cite{vanKempen02}) are shown as dashed lines. The shaded regions mark the range of frequencies consistent with an error of $0.2\,a_B$ in the scattering lengths and an uncertainty of 10\% (as upper bound) in the lattice depth, where both errors have been added.}
	\label{fig:Paper_OscFreq}
\end{figure}

For these two initial two-particle states \ket{0,0} and \ket{-1,-1}, spin oscillations have been observed for various magnetic fields, corresponding to different detunings. The measured oscillation frequency, plotted in Fig.~\ref{fig:Paper_OscFreq}, are well fitted by the expected behaviour of the effective Rabi frequency $\Omega^\prime(B)$ with varying detuning (solid lines in Fig.~\ref{fig:Paper_OscFreq}). 
As the magnetic field dependence $\delta(B^2)$ is known, this fit allows to extract the bare coupling strength $\Omega_{if}$ and zero-field detuning $\delta_0$, related to known combinations of the characteristic scattering lengths $a_F$ in the way explained above. The parameters for the different processes determined by the fits are summarized in Table \ref{tab:EffectiveRabiFrequencies}.
\begin{table}[htbp]
	\begin{center}
		\begin{tabular}{r@{$\leftrightarrow$}l|c|c|c}
			\multicolumn{2}{c|}{Process} & $\Omega_{if}/2\,\pi$ & $\delta_{0}/2\,\pi$ & $\Omega_{if}^\prime(B=0)/2\,\pi$\\ \hline \hline
			$\ket{0,0}$ & $\ket{+1,-1}$ &  243(2)\,Hz & 39(2)\,Hz & 247(3)\,Hz\\ \hline
			$\ket{-1,-1}$ & $\ket{0,-2}$ & 212(1)\,Hz & 26(1)\,Hz& 213(3)\,Hz 
		\end{tabular}
	\end{center}
	\caption{Summary of measured coupling strength $\Omega_{if}$, zero-field detuning $\delta_0$ and effective Rabi frequency $\Omega^\prime_{if} = \sqrt{\delta_0^2 + \Omega_{if}^2}$ at $40\,E_r$ lattice depth.}
	\label{tab:EffectiveRabiFrequencies}
\end{table}
These measurements represent a high precision test of spin-dependent scattering properties and can be used in particular to test potentials for ultracold collisions for two Rb atoms. Knowledge of these potentials allows to compute the scattering lengths $a_F$, from which theoretical values for $\Omega_{if}$ and $\delta_0$ can be calculated for our trapping conditions and can be directly compared to the experiment. We find a general agreement in Fig.~\ref{fig:Paper_OscFreq} between the measured oscillation frequencies and the calculated ones, based on the prediction of \cite{vanKempen02} for the $a_F$. However, a discrepancy on the order of 20\% remains. More details on this comparison will be given in a forthcoming publication \cite{Widera05pre}.

\begin{figure}[htbp]
	\begin{center}
		\includegraphics[scale=0.7]{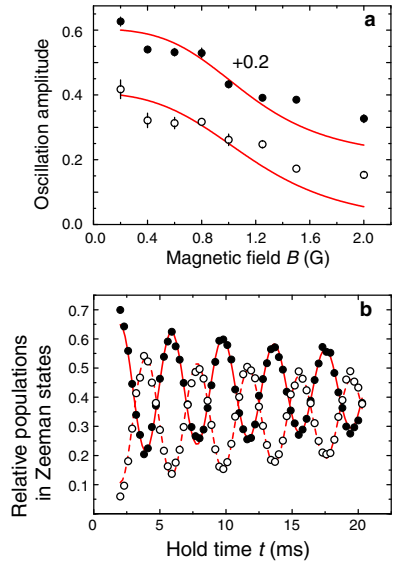}
	\end{center}
	\caption{(a) Oscillation amplitude vs.~magnetic field for the processes $\ket{00} \leftrightarrow \ket{+1,-1}$ ($\medcirc$) and $\ket{-1,-1}\leftrightarrow \ket{0,-2}$ ($\medbullet$). The lines are fits to the expected behavior of the Rabi amplitude, where only the fraction of atoms $n$ in sites with a filling of two remains as free parameter. The upper curve has been offset by 0.2. (b) High contrast spin oscillations between $\ket{0,0}$ ($\medcirc$) and $\ket{+1,-1}$ ($\medbullet$) at 0.6\,G and $40\,E_r$. Here, atoms in sites with unity filling are selectively discarded from the measurement (see text).}
	\label{fig:Paper_OscRemoved_3V1}
\end{figure}

The visibility of the oscillations shown in Fig.~\ref{fig:Paper_Oscillations}a is smaller than the one expected from the Rabi model alone for this detuning. 
This is due to the fact that only those atoms with two atoms per site contribute to the spin dynamics, whereas the relative populations are normalized with respect to the total atom number. 
Hence the visibility is artificially decreased, which we accounted for by the factor $n$ introduced above. With the extracted values of $\Omega_{if}$ and $\delta_{if}$ we fit $n$ to the amplitude of the spin oscillation (see solid line in Fig.~\ref{fig:Paper_OscRemoved_3V1}a). The mean measured fraction resulting from the fit is $n=0.42(3)$ close to the value $n \approx 50\%$ calculated for the ground state.

Further evidence comes from a separate experiment, where we show that the contrast can be enhanced by selectively discarding atoms from the measurement which are in sites with unity filling. This is accomplished by first evolving the system for one half period of spin-oscillation, at which the occupation of the $\ket{+1,-1}$-state is maximal in sites with two atoms. Atoms in the $m_f = 0$ state, predominantly in singly occupied sites, are then transferred into the $f=1$ hyperfine manifold by the application of a $6\,\mu$s microwave $\pi$-pulse. As the probe light of the absorption imaging is resonant with the $f=2 \rightarrow f^\prime=3$ optical transition only, those atoms remain undetected in the subsequent spin-evolution. The resulting high contrast oscillations are shown in Fig.~\ref{fig:Paper_OscRemoved_3V1}b for the case of $B=0.6$\,G.  
We measure an increase of the initial amplitude from 0.31(3) without the microwave to 0.56(4). The difference to the theoretically expected value of 0.8 can be explained by a small fraction of atoms in $m_f=\pm 2$ (approximately 17\%) \cite{Note04}. 

\begin{figure}[htbp]
	\begin{center}
		\includegraphics[scale=0.85]{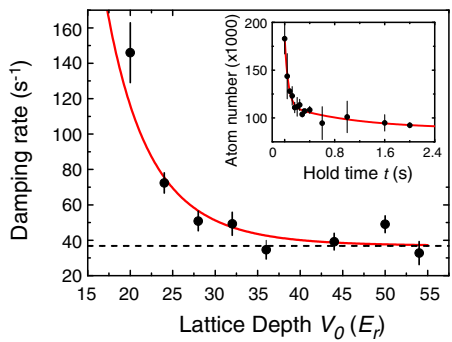}
	\end{center}
	\caption{Damping rate of spin oscillations vs.~lattice depth  for $\ket{0,0} \leftrightarrow \ket{+1,-1}$. The solid line is a fit to $\alpha J/h+\gamma_0$ with $\gamma_0=37(3)\,\mbox{s}^{-1}$ (illustrated as dashed line). The inset shows atom loss vs.~hold time at a 10\,G magnetic field, where spin dynamics is suppressed. The fast loss rate of $\gamma_1 \approx 35(5)\,\mbox{s}^{-1}$ coincides with the offset $\gamma_0$ of the damping rate even at high lattice depths.}
	\label{fig:Paper_Damping}
\end{figure}

The picture of isolated atom pairs coherently oscillating between two-particle states does not allow to understand the damping of these oscillations, clearly seen in Fig.~\ref{fig:Paper_Oscillations}. 
To investigate the damping mechanism, we record the decay rate as a function of lattice depth at $B=0.6$\,G (see Fig.~\ref{fig:Paper_Damping}). Here the depth has been changed between $20\,E_r$ and $54\,E_r$ in both horizontal lattice axes.
For high lattice depths the damping rate levels at a finite value, whereas it increases for decreasing lattice depths. For depths smaller than $20\,E_r$ the coherent oscillations could not be recorded any more. We interpret the increase as a signature of the onset of tunneling in the lattice. 
A fit to the damping rate of the form $\alpha \, J/h + \gamma_0$ returns a proportionality constant $\alpha = 9(1)$ and a constant offset $\gamma_0=37(3)\,\mbox{s}^{-1}$ (see solid line in Fig.~\ref{fig:Paper_Damping} for the fit and dashed line for the offset).
This offset can be explained by the atom loss rate in the lattice, as shown in the inset of Fig.~\ref{fig:Paper_Damping}. For this experiment, the sample was prepared in the $\ket{0,0}$ state and spin dynamics was suppressed by a magnetic field around $10$\,G. The initial loss rate $\gamma_1=35(5)\,\mbox{s}^{-1}$, attributed to two-body inelastic processes with a measured two-body loss coefficient $K_2 = (8.8 \pm 1.5)\times 10^{-14}\,\mbox{cm}^3/\mbox{s}$, agrees with the offset in the damping within the quoted uncertainties.
We have also detected spin dynamics in the $f=1$ hyperfine manifold between the two-particle states $\ket{0,0} \leftrightarrow \ket{+1,-1}$ and find a coupling strength roughly one order of magnitude smaller than in $f=2$ \cite{Widera05pre}, in agreement with theoretical predictions.

In summary, we have observed coherent spin dynamics between two-particle states in the upper hyperfine ground state of \Rb\ due to spin changing collisions. The observation of high contrast Rabi-type oscillations make this system a promising starting point for quantum information purposes. 
In this work we have demonstrated a method to create an array of entangled atom pairs similar to the long-lived Bell pairs produced in ion traps \cite{Roos04}, which constitutes a first step towards the creation of pair-correlated atomic beams as proposed in \cite{Pu00,Duan00}.
Another intriguing question is the evolution of quantum correlations upon melting the Mott-insulator. A possible outcome would be a non-local condensate of Bell-like pairs delocalized over the entire cloud, which could be detected through counting statistics \cite{ParedesPrivate}.

We would like to thank Servaas Kokkelmans for helpful discussions. This work was supported by the DFG, the European Union (OLAQUI) and the AFOSR. FG acknowledges support from a Marie-Curie fellowship.

\end{document}